\documentclass[aps,prl,twocolumn,superscriptaddress,notitlepage,nofootinbib]{revtex4-1}
\usepackage{makecell}
\usepackage{amsmath}
\usepackage{bm}
\usepackage{graphicx}
\usepackage{color,soul}
\usepackage{float}
\usepackage{multirow}
\usepackage[inline]{enumitem}
\usepackage{cancel}
\usepackage[normalem]{ulem}
\usepackage[breaklinks,colorlinks,urlcolor=blue,citecolor=citcolor,linkcolor=lcolor]{hyperref}
\definecolor{lcolor}{rgb}{0.5,0,0}
\definecolor{citcolor}{rgb}{0,0.3,0.0}
\definecolor{ao(english)}{rgb}{0.0, 0.5, 0.0}

\newcommand{\vect}[1]{\boldsymbol{#1}_{\perp}}
\newcommand{\pgammat}{\boldsymbol{p_{\gamma\perp}}}
\newcommand{\lt}{\boldsymbol{l_{\perp}}}

\newcommand{\der}{\mathrm{d}}

\newcommand{\ltU}[1]{\boldsymbol{l}_{\perp}^{#1}}

\newcommand{\Lt}{\vect{L}}
\newcommand{\LtL}[1]{\boldsymbol{L}_{\perp #1}}
\newcommand{\LtU}[1]{\boldsymbol{L}_{\perp}^{#1}}

\newcommand{\ltC}{\vect{l}'}
\newcommand{\ltCL}[1]{\boldsymbol{l}'_{\perp #1}}
\newcommand{\ltCU}[1]{\boldsymbol{l}_{\perp}^{'#1}}

\newcommand{\pgammatL}[1]{\boldsymbol{p}_{\gamma\perp #1}}
\newcommand{\pgammatU}[1]{\boldsymbol{p}_{\gamma\perp}^{#1}}

\def\bea#1\eea{\begin{align}#1\end{align}}
\newcommand{\nn}{\nonumber\\}
\newcommand{\bef}{\begin{figure}[h!tb]\centering}
\newcommand{\eef}{\end{figure}}

\newcommand{\yt}{\boldsymbol{y}_{\perp}}
\newcommand{\ytC}{\boldsymbol{y'}_{\perp}}
\newcommand{\ytone}{\boldsymbol{y}_{1\perp}}
\newcommand{\yttwo}{\boldsymbol{y}_{2\perp}}

\allowdisplaybreaks

\begin{document}
\title{Correspondence between Color Glass Condensate and High-Twist Formalism}

\author{Yu Fu}
\affiliation{Key Laboratory of Quark and Lepton Physics (MOE) \& Institute of Particle Physics, Central China Normal University, Wuhan 430079, China}
\affiliation{Department of Physics, Duke University,
Durham, North Carolina 27708, USA}

\author{Zhong-Bo Kang}
\affiliation{Department of Physics and Astronomy, University of California, Los Angeles, California 90095, USA}
\affiliation{Mani L. Bhaumik Institute for Theoretical Physics, University of California, Los Angeles, California 90095, USA}
\affiliation{Center for Frontiers in Nuclear Science, Stony Brook University, Stony Brook, New York 11794, USA}

\author{Farid Salazar}
\affiliation{Nuclear Science Division, Lawrence Berkeley National Laboratory, Berkeley, California 94720, USA}
\affiliation{Physics Department, University of California, Berkeley, California 94720, USA}
\affiliation{Department of Physics and Astronomy, University of California, Los Angeles, California 90095, USA}
\affiliation{Mani L. Bhaumik Institute for Theoretical Physics, University of California, Los Angeles, California 90095, USA}

\author{Xin-Nian Wang}
\affiliation{Key Laboratory of Quark and Lepton Physics (MOE) \& Institute of Particle Physics, Central China Normal University, Wuhan 430079, China}
\affiliation{Nuclear Science Division, Lawrence Berkeley National Laboratory, Berkeley, California 94720, USA}

\author{Hongxi Xing}
\affiliation{Key Laboratory of Atomic and Subatomic Structure and Quantum Control (MOE), Guangdong Basic Research Center of Excellence for Structure and Fundamental Interactions of Matter, Institute of Quantum Matter, South China Normal University, Guangzhou 510006, China}
\affiliation{Guangdong-Hong Kong Joint Laboratory of Quantum Matter, Guangdong Provincial Key Laboratory of Nuclear Science, Southern Nuclear Science Computing Center, South China Normal University, Guangzhou 510006, China }
\affiliation{Southern Center for Nuclear-Science Theory (SCNT), Institute of Modern Physics, Chinese Academy of Sciences, Huizhou 516000, China}

\begin{abstract}
The Color Glass Condensate (CGC) effective theory and the collinear factorization at high-twist (HT) are two well-known frameworks describing perturbative QCD multiple scatterings in nuclear media. It has long been recognized that these two formalisms have their own domain of validity in different kinematics regions. Taking direct photon production in proton-nucleus collisions as an example, we clarify for the first time the relation between CGC and HT at the level of a physical observable. We show that the CGC formalism beyond shock-wave approximation, and with the Landau-Pomeranchuk-Migdal interference effect is consistent with the HT formalism in the transition region where they overlap. Such a unified picture paves the way for mapping out the phase diagram of parton density in nuclear medium from dilute to dense region. 
\end{abstract}

\maketitle

\noindent{\it Introduction.--} In high-energy scatterings involving heavy nuclei, many interesting nuclear dependent phenomena have been observed~\cite{PHENIX:2004nzn,PHENIX:2019gix,ALICE:2021est,PHOBOS:2004fsu,BRAHMS:2004xry,ATLAS:2016xpn,PHENIX:2017caf,LHCb:2021vww,LHCb:2022rlh,ALICE:2018vuu,STAR:2021fgw,Braidot:2010zh,STAR:2006dgg,PHENIX:2011puq}. 
The essential ingredient for understanding novel nuclear dependence in different collision systems is the description of multiple parton scattering inside the nuclei. It is thus critical to elucidate these multiple scatterings in perturbative QCD (pQCD) in different kinematic regimes~\cite{Boer:2011fh,Accardi:2012qut,Aschenauer:2017jsk,AbdulKhalek:2021gbh} of the nuclear medium.

\begin{figure}[htbp]
	\centering	\includegraphics[width=0.5\textwidth]{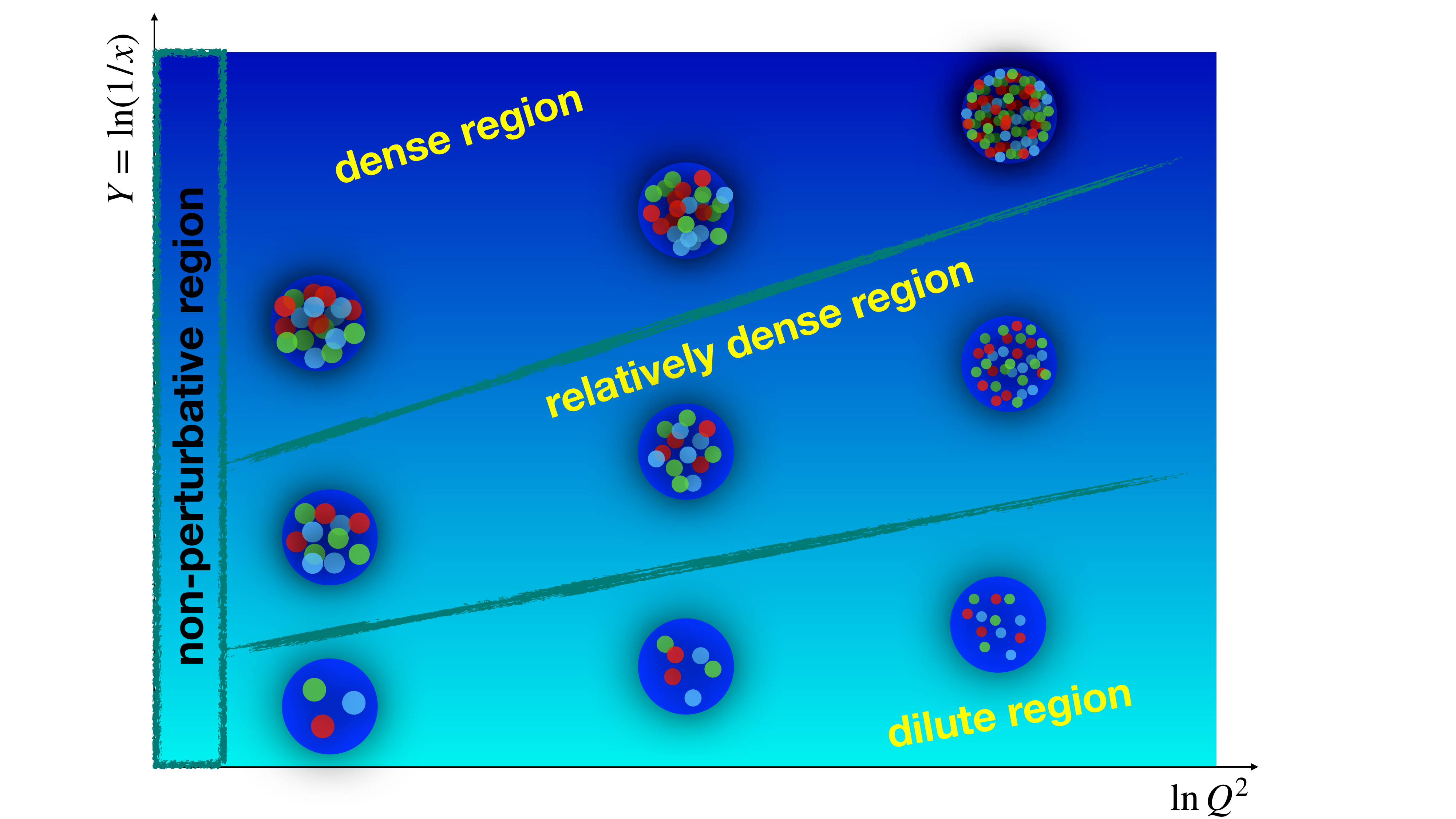}
\caption{Phase diagram of parton density in nuclear medium in terms of momentum fraction $x$ and probing scale $Q$.}
	\label{fig-density}
\end{figure}

The Color Glass Condensate (CGC) effective theory~\cite{McLerran:1993ni,McLerran:1993ka,McLerran:1994vd,Ayala:1995kg,Ayala:1995hx,Gelis:2010nm,Kovchegov:2012mbw} and the collinear factorization at high-twist (HT)~\cite{Ellis:1982wd,Qiu:1990xxa,Qiu:1990xy} are two well-known theoretical frameworks describing QCD multiple scatterings in nuclear media. They have been extensively used to describe the phase diagram of parton density in nucleon/nuclei as shown schematically in Fig. \ref{fig-density}, as a function of parton momentum fraction $x$ and the associated hard scale $Q$. In the dilute region where $x\sim \mathcal{O}(1)$, the corresponding pQCD collinear factorized formalism at leading twist~\cite{Collins:1989gx} has been very successful and set as a benchmark theory for high-energy physics. In the relatively dense region where $x \lesssim \mathcal{O}(1)$, the high-twist expansion approach based on the generalized QCD collinear factorization theorem~\cite{Qiu:1990xxa,Qiu:1990xy} provides a robust framework to describe multiple scatterings in nuclear medium order by order, which are essentially power corrections to the leading twist cross-section. Such an approach has been successfully applied to calculate the incoherent multiple scattering at the next-to-leading power~\cite{Kang:2013ufa,Kang:2014hha}, and to the study of jet quenching in cold nuclei~\cite{Guo:2000nz,Wang:2001ifa}. In the high energy limit, $x \sim 1/\sqrt{s} \to 0$, the gluon density grows rapidly resulting in a high gluon occupation number. It is expected that the gluon density is tamed by non-linear QCD effects at sufficiently small-$x$ \cite{Gribov:1984tu,Mueller:1985wy}. The CGC provides an effective description of this saturated regime, with many experimental consequences \cite{Kharzeev:2004yx,Marquet:2007vb,Lappi:2012nh,Albacete:2018ruq,Zheng:2014vka,Lappi:2013zma,JalilianMarian:2012bd,Ducloue:2017kkq,Ducloue:2015gfa,Shi:2021hwx,Tong:2022zwp,Benic:2022ixp,Al-Mashad:2022zbq,Liu:2022ijp,Liu:2023aqb,Caucal:2023fsf,Morreale:2021pnn}.

It has long been recognized that the HT and CGC approaches have drastic differences. One of the main differences is the QCD factorization theorems they rely on. The HT approach follows the generalized QCD collinear factorization, in which the medium property is encoded in the multi-parton quantum correlation functions satisfying the DGLAP-type evolution \cite{Kang:2013raa,Kang:2014ela,Kang:2016ron}. The CGC, on the other hand, follows transverse momentum-dependent factorization at small-$x$, and the corresponding medium properties are encoded in correlators of light-like Wilson lines, which satisfy the Jalilian-Marian-Iancu-McLerran-Weigert-Leonidov-Kovner/Balitsky-Kovchegov nonlinear evolution \cite{Balitsky:1995ub,Kovchegov:1999ua,Jalilian-Marian:1997qno,Jalilian-Marian:1997jhx,Jalilian-Marian:1997ubg,Kovner:2000pt,Iancu:2000hn,Iancu:2001ad,Ferreiro:2001qy}. In terms of multiple scattering, extra soft rescatterings are considered order by order in a power series in addition to the hard scattering in HT approach, while in the CGC analysis, all scatterings are treated on the same footing and within the eikonal approximation which allows for their exponentiation into the light-like Wilson line. 

Despite the many successes of the HT and CGC formalisms, they were limited to their own domain of validity. It is believed that they have to agree with each other in the overlap region where both are applicable, simply because of the universality of the medium property they probe. There have been tremendous efforts to show the correspondence between CGC and QCD collinear factorization, aiming to extend the applicability of CGC from small-$x$ (dense) to large-$x$ (dilute) region with particular emphasis on the sub-eikonal corrections to the parton propagators \cite{Altinoluk:2014oxa,Altinoluk:2015gia,Altinoluk:2015xuy,Agostini:2019avp,Agostini:2019hkj,Agostini:2022ctk,Altinoluk:2020oyd,Altinoluk:2021lvu, Chirilli:2018kkw,Chirilli:2021lif}, the rapidity evolution of unintegrated gluon distributions \cite{Balitsky:2015qba,Balitsky:2016dgz,Balitsky:2017flc,Balitsky:2017gis,Balitsky:2019ayf}, as well as new semi-classical approaches \cite{Boussarie:2020fpb,Boussarie:2021wkn,Boussarie:2023xun,Jalilian-Marian:2017ttv,Jalilian-Marian:2018iui,Jalilian-Marian:2019kaf,Kovner:2023vsy}\footnote{Sub-eikonal corrections are also necessary to describe the physics of spin at small-$x$ \cite{Kovchegov:2015pbl,Kovchegov:2016weo,Kovchegov:2017lsr,Kovchegov:2018znm,Kovchegov:2020hgb,Cougoulic:2022gbk,Borden:2023ugd,Adamiak:2023okq,Li:2023tlw}.}. However, no consensus has yet been reached on the relations between HT and CGC and the identification of transition mechanisms from dilute to dense regions.

In this letter, we clarify for the first time the correspondence between HT and CGC formalisms at the level of a physical observable. In particular, taking direct photon production in $pA$ collisions as an example, we present a systematic treatment of the nuclear enhanced initial- and final-state double scatterings, as well as their interference. We prove the consistency between HT and CGC by going beyond the shock wave approximation and including the Landau-Pomeranchuk-Migdal (LPM) interference effect \cite{Landau:1955,Migdal:1956tc}. We argue that the generalization of such an approach to all hard scattering processes is straightforward. Therefore, our results provide a unified picture of dilute-dense dynamics in nuclear media. It paves the way to mapping out the phase diagram of atomic nuclei in terms of parton density as shown in Fig. \ref{fig-density}, and to understanding the underlying multiple scattering mechanisms.

\vspace{0.5cm}
\noindent{\it Dilute versus dense regions.}- In order to show explicitly the correspondence between the CGC and the generalized collinear factorization formalism, we take direct photon production in $pA$ collisions as an example, $p(P_p^-)+A(P_A^+) \to \gamma(p_{\gamma})+X$,
where $P_p^-$, $p_{\gamma}$ and $P_A^+$ are, respectively, the momentum for the incoming proton, the observed photon and the averaged momentum per nucleon inside the nucleus. The direct photon production has a unique advantage to test QCD multiple scattering effects due to the absence of strong interaction between the photon and the nuclear medium. We focus on the interactions between quarks from the proton and gluons from the nucleus. The extension to other channels and processes can be performed in a similar fashion.     

\begin{figure}[hbtp]
	\centering	\includegraphics[width=0.48\textwidth]{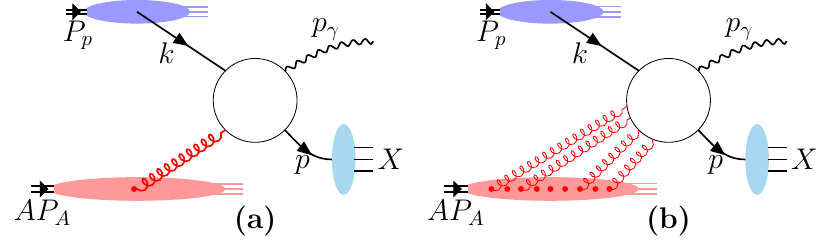}
\caption{Schematic diagrams for single (a) and multiple (b) scatterings for direct photon production in $pA$ collisions, the circles represent quark-gluon hard interaction.}
	\label{fig-exp}
\end{figure}

In collisions involving a large nucleus with mass number $A$, effects of multiple scatterings can be enhanced by powers of the nuclear size, $L_A^-\sim A^{1/3}$,  thus becoming important.
Such multiple scatterings can be described by the generalized factorization formalism order by order as illustrated in Fig. \ref{fig-exp}, i.e. $\der \sigma = \der \sigma^{\rm LT} + \der \sigma^{\rm T4}+\dots$, where LT and T4 stand for leading twist and twist-4, respectively. In the dilute region, the cross-section is dominated by partonic processes of single scattering, as shown in Fig. \ref{fig-exp}(a). The standard LT factorization yields~\cite{Collins:1989gx}
\bea
E_{\gamma}\frac{\der^3\sigma^{\rm LT}}{\der^3 \boldsymbol{p_{\gamma}}} =& f_{q/p}(x_q)\otimes xf_{g/A}(x)\otimes H^{(2)}_{q+g\to \gamma + q}\,,
\label{eq-t2}
\eea
where $\otimes$ stands for the convolution of the LT quark $f_{q/p}$ and gluon $f_{g/A}$ distribution function and the hard partonic part
$H^{(2)}_{q+g\to \gamma+q} = (e_q^2\alpha_{em}\alpha_{s}/N_c)  \xi^2[1+(1-\xi)^2]/p_{\gamma\perp}^4$ for single scattering $q+g\to \gamma+q$ \cite{Owens:1986mp}, where $\xi = p_{\gamma}^-/(x_qP_p^-)$. The summation over the quark flavor index is implicit throughout this Letter.

The leading nuclear corrections beyond the single scattering picture can be formulated within the framework of generalized  factorization. Previous studies of direct photon production in $pA$ collisions have focused on contributions from initial-state double scattering \cite{Guo:1995zk,Kang:2013ufa}, which are dominant in the large-$x$ region. In this study, using the generalized factorization, we go beyond the large-$x$ region and calculate for the first time the complete result including both initial and final state double scatterings, as well as their interference. The final results can be written schematically as,
\bea
E_{\gamma}\frac{\der^3\sigma^{\rm T4}}{\der^3 \boldsymbol{p_{\gamma}}} =& f_{q/p}\otimes \mathcal{D}_{X}T_{gg} \otimes H^{(4)}_{q+gg\to \gamma+q}\,,
\label{eq-t4}
\eea
for each cut diagram, where $H^{(4)}$ is the corresponding hard function at twist-4, $T_{gg}$'s are twist-4 gluon-gluon correlation functions in the nucleus, for example,
\begin{eqnarray}
    T_{gg}(\{x_i\}) \!\!&=& \!\!\!\!\int_{\{y_i^-\}} e^{iP_A^+\left[x_1y^- + x_2 (y_1^- - y_2^-) + x_3  y_2^-\right]} \theta(y^--y_1^-)
\nonumber \\ 
    && \hspace{-0.66in} \times \theta(-y_2^-)
    \langle P_A|F_{\alpha}^+(y_2^-)F^{\beta +}(0)F_{\beta}^+(y^-)F^{+\alpha}(y_1^-)|P_A\rangle,
    \label{eq-Tgg}
\end{eqnarray}
with $\int_{\{y_i^-\}} \equiv [1/(4\pi^2P_A^+)]\int dy^- dy_1^-  dy_2^-$. The short-hand notation $\mathcal{D}_{X}$ stands for the linear combination of derivatives $\partial/\partial x_i$ and $\partial^2/(\partial x_i \partial x_j)$ that act on $ T_{gg}(\{x_i\})\equiv T_{gg}(x_1,x_2,x_3)$, and we call it non-derivative contribution when $\mathcal{D}_{X}=1$. Unlike the LT parton distribution functions, which possess a probability density interpretation, the twist-4 matrix elements characterize the quantum parton-parton correlations inside the nucleus. The detailed derivation will be given in a forthcoming companion paper \cite{Fu:2023xxx}, and we provide the complete expressions for Eq.~(\ref{eq-t4}) in the supplemental material.

In the extremely dense region ($x\to 0$) as shown in Fig. \ref{fig-density}, the gluon occupation number becomes large, and the probe coherently interacts with the entire nucleus, where the coherence length $\lambda_c\sim 1/xP_A^+ \gg L_A^-$. In this regime, one must resum coherent multiple scatterings to all orders. In the CGC effective theory, within the hybrid factorization formalism \cite{Gelis:2002ki}, the cross-section can be written as 
\bea
    E_{\gamma} \frac{\der^3\sigma^{\mathrm{CGC}}}{\der^3 \boldsymbol{p_{\gamma}}} = &   \frac{e_q^2 \alpha_{em}}{2\pi^2} \xi^2[1+(1-\xi)^2] \otimes f_{q/p}(x_q)\nn
    &\otimes \int \der^2\lt\frac{\lt^2F(x,\lt)}{(\xi \lt-\pgammat)^2\pgammat^2}\,,
    \label{eq-cgc}
    \eea
where the dipole distribution is defined as
    \bea
    F(x,\lt)=&\int\frac{\der^2\yt}{2\pi}\int\frac{\der^2\ytC}{2\pi}e^{-i\lt\cdot(\yt-\ytC)}
    \nn
    &\times\frac{1}{N_c}\langle {\rm Tr}[V^\dagger(\ytC)V(\yt)]\rangle\,,
    \label{eq-dipole}
\eea
and $V(\yt)=\mathcal{P}\left[\exp\left(ig \int \der y^- A^+(\yt,y^-) \right) \right]$ stands for the light-like Wilson line, encoding the multiple eikonal scattering of the projectile quark with the nucleus. Here, $\langle \dots \rangle_x$ stands for the average over different classical color charge configurations in the CGC. The cross-section in Eq.~(\ref{eq-cgc}) has a collinear divergence at $\pgammat = \xi\lt$,  which can be regularized by the redefinition of the photon fragmentation function \cite{Koller:1978kq,Laermann:1982jr}.\\

\noindent{\it Mismatch between HT and power expansion of CGC.}
- In this Letter, we are aiming to find the link between the CGC and HT beyond the small-$x$ limit. Such a kinematic region can be realized when $p_{\gamma\perp}$ is larger than the saturation scale $Q_s\sim \langle l_{\perp}\rangle$, which is the typical transverse momentum in the multiple parton scattering. To see the connection to the high-twist formalism, we perform Taylor (or collinear) expansion  of the CGC result in powers of $Q_s^2/\pgammat^2$ for $p_{\gamma\perp} > Q_s$. We also use the following relations between the collinear gluon distributions and the moments of dipole distribution,
\begin{align}
    \lim_{x\to 0}xf_{g/A}(x) =& \frac{N_c}{2\pi^2\alpha_s}\int \der^2 \lt\lt^2F(x,\lt)\,, \\
    \lim_{x\to 0}T_{gg}(x,0,0) =& \frac{N_c^2}{2(2\pi)^4\alpha_s^2}\int \der^2 \lt \lt^4F(x,\lt)\,.
    \label{eq:4-moment-dipole}
\end{align}
The first relation above has been long established in Refs.~\cite{Baier:2004tj,Kang:2012vm}, whereas the second equation is derived for the first time and will be presented in the companion paper \cite{Fu:2023xxx}. 
The CGC result in Eq.\,\eqref{eq-cgc} after the collinear expansion in small-$x$ limit becomes,
\begin{eqnarray}
    E_{\gamma}\frac{\der^3\sigma^{\rm{CGC}}}{\der^3 \boldsymbol{p_{\gamma}}} 
    &=& f_{q/p}(x_q) \otimes H^{(2)}_{q+g\to \gamma + q}
    \nonumber    \\     
    & & \hspace{-1.0in}\otimes \left[x f_{g/A}(x)+\frac{(2\pi)^2\alpha_s}{N_c}\frac{4 \xi^2}{p_{\gamma\perp}^2}T_{gg}(x,0,0)+\cdots\right]_{x\to 0 }.
    \label{eq-CGC_exp}
\end{eqnarray}
One immediately sees that the first term matches the LT result in Eq.~(\ref{eq-t2}) in the limit of $x\to 0$, where the longitudinal phase $e^{ixP_A^+y^-}$ in $f_{g/A}$ can be neglected. Such matching between CGC and LT results has been realized in other processes \cite{Gelis:2003vh,Kang:2012vm,Benic:2016uku}. However, the matching to HT formalism has never been established. The second term in Eq.~(\ref{eq-CGC_exp}) reproduces the non-derivative term in twist-4 result in Eq.~(\ref{eq-t4}) if one neglects all the longitudinal phases in Eq.~(\ref{eq-Tgg}) and assumes all the twist-4 distributions reduce to the universal object at small $x$ in Eq.~\eqref{eq:4-moment-dipole}. Since the derivative terms in Eq.~(\ref{eq-t4}) also arise from the longitudinal phases that get entangled with the collinear expansion, they are one of the primary reasons for the mismatch between the CGC and HT formalism.

\vspace{0.5cm}
\noindent {\it Sub-eikonal phases and LPM interference}.
- A rigorous proof of the matching between CGC and HT factorization at finite $x$ is nontrivial, and has recently triggered various efforts \cite{Altinoluk:2014oxa,Altinoluk:2015gia,Altinoluk:2015xuy,Agostini:2019avp,Agostini:2019hkj,Agostini:2022ctk,Altinoluk:2020oyd,Altinoluk:2021lvu,Chirilli:2018kkw,Chirilli:2021lif,Balitsky:2015qba,Balitsky:2016dgz,Balitsky:2017flc,Balitsky:2017gis,Balitsky:2019ayf,Jalilian-Marian:2017ttv,Jalilian-Marian:2018iui,Jalilian-Marian:2019kaf,Boussarie:2020fpb,Boussarie:2021wkn,Boussarie:2023xun}. In this Letter, we reveal for the first time two key ingredients for the matching: sub-eikonal phases and the LPM effect, in proving the exact correspondence at twist-4 level in finite-$x$ region. 

In the CGC, the eikonal scatterings between the fast projectile and the nucleus' small-$x$ background field can be resummed into an effective vertex, known as the shock wave approximation, allowing one to write down the dipole distribution in a compact form shown in Eq.~(\ref{eq-dipole}). The price paid in such a compact expression is the neglect of the information encoded in the longitudinal phase factors, which is essential at finite-$x$. Therefore, we must first bring back the longitudinal sub-eikonal phases to restore the information associated with the non-zero longitudinal momentum transfer. These sub-eikonal phases can not be easily exponentiated to all orders, we thus examine the corresponding twist contributions from the CGC effective vertex by expanding the light-like Wilson line in powers of gauge field $A^+$. At leading order in the expansion, the interacting vertex between the quark projectile and the nuclear medium becomes: 
\begin{align}
    \Gamma_{q}(l) =  &(2\pi)\delta(l^-) \gamma^-  \int_{y} e^{-i \lt \cdot \yt}  e^{i l^+y^-} ig A^{+}(y^-,\yt)\,, \nonumber
\end{align}
where $l$ denotes the momentum transfer from the medium to the quark, and we introduced the short-hand $\int_y \equiv   \int \der^2 \yt \int \der y^-$. Armed with this vertex we find that the single scattering contribution reads
\begin{align}
    E_{\gamma}\frac{\der^3\sigma_{\mathrm{S}}^{{\rm CGC_{sub}}}}{\der^3 \boldsymbol{p_{\gamma}}} = & f_{q/p}(x_q) \otimes \int_{y,y'} \!\!\!\! \mathcal{H}_{\rm S}  \langle \mathrm{Tr}\left[A^+(y)A^+(y') \right]\rangle.
\end{align}
The explicit expression for the perturbative factor $\mathcal{H}_{\rm S}$ is given in the supplemental material.
The leading term in the expansion of $\mathcal{H}_{\rm S}$ in inverse powers of $p_{\gamma\perp}^2$ is
\begin{align}
    &\mathcal{H}_{\rm S}(p_\gamma,y,y') = \frac{2 }{\pi} H^{(2)}_{q+g\to \gamma+q}  e^{i x P_A^+ (y^- - y'^-)}    \nonumber \\
    &\times \delta^{(2)}(\yt-\vect{y'}) (\partial_{\yt} \cdot \partial_{\ytC}) + \mathcal{O}(1/\pgammat^6)    \,.
     \label{eq-hexp1}
\end{align}
The derivatives convert the gauge field $A^+$ into the field strength tensor $ (\partial_{\yt} \cdot \partial_{\ytC}) A^+ A^+ \to F_{\perp}^+ \cdot F_{\perp}^+$, which eventually leads to the exact matching to the standard leading twist collinear factorization result shown in Eq.\,(\ref{eq-t2}), including the longitudinal phase factor $e^{ixP_A^+y^-}$. As it is customary, to make this identification, we employed the correspondence between the CGC average and the quantum average~\cite{Dominguez:2011wm}, $\langle \mathcal{O} \rangle = \langle P_A| \mathcal{O} |P_A\rangle/\langle P_A|P_A\rangle$ with $\langle P_A|P_A'\rangle = 2 P_A^+ \delta^{(3)}(P_A-P_A')$. 

\begin{figure}[H]
    \centering
    \includegraphics[width=0.48\textwidth]{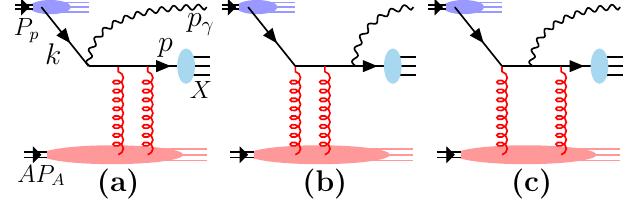}
    \caption{Double scattering diagrams for direct photon production in $pA$ collisions.}
    \label{fig-double}
\end{figure}

In connecting to the complete twist-4 contribution, we need to consider the expansion of the light-like Wilson line in the CGC up to three gluon fields at amplitude level, corresponding to triple scatterings. In the following, we take double scattering as an example to show explicitly the matching. The extension to single-triple interference process can be carried out in the same fashion and will be detailed in Ref.~\cite{Fu:2023xxx}. 

As shown in Fig.~\ref{fig-double}, there are three diagrams that contribute to double scatterings at the amplitude level. In the eikonal approximation employed in the CGC, emissions between scatterings are omitted \cite{Gelis:2002ki}, thus only diagrams $(a,b)$ contribute, corresponding to initial- and final-state double scattering, respectively. However, by keeping track of the longitudinal sub-eikonal phases, one observes that the computation of diagram $(c)$ yields two different contributions that only differ by an overall phase factor. In terms of the formation time of the radiated photon, $\tau_{\gamma} = 2x_q P_p^-\xi(1-\xi)/(\pgammat - \xi \lt)^2$, this phase difference can be expressed as $1-e^{i \Delta y^- /\tau_{\gamma}}$, where $\Delta y^-$ is the distance between scattering locations. It is clear that in high energy limit $P_p^-\to \infty$, there is a perfect destructive interference thus this diagram vanishes. This cancellation displays the characteristic LPM effect \cite{Landau:1955,Migdal:1956tc}, revealing the fact that when the photon formation time is larger than the two scattering centers $\tau_{\gamma} \gg \Delta y^-$, the photon becomes coherent and can not resolve the two different scatterings.  However, in the finite-$x$ region, the phases do not cancel each other completely and therefore there remains a net contribution at the twist-4 level, which is required to establish the matching with the HT formalism.  The LPM effect has been studied extensively in the context of parton energy loss \cite{Baier:1996kr,Baier:1996sk,Zakharov:1996fv,Wiedemann:2000za,Mehtar-Tani:2011hma,Sievert:2018imd,Dominguez:2019ges,Arleo:2020hat,Caron-Huot:2010qjx}, but this is the first time it is emphasized within the context of matching CGC and HT.  
 
Similar types of diagrams are also neglected in the single-triple interference processes from CGC expansion. We emphasize again that such types of diagrams are non-negligible in the finite-$x$ region due to the LPM effect, which is another important ingredient in the exact matching between CGC and HT. Including these two missing ingredients, we obtain the following result 
\begin{align}
    & E_{\gamma}\frac{\der^3\sigma_{\rm D}^{{\rm CGC_{sub}}}}{\der^3 \boldsymbol{p_{\gamma}}} 
    = f_{q/p}(x_q) \otimes \int_{\substack{y,y'\\
    y_1,y_2}} \!\!\!\!  \ \ \Theta (y,y',y_1,y_2) \nonumber \\ 
    & \ \ \ \ \ \times  \mathcal{H}_{\rm D} \langle \mathrm{Tr}\left[A^+(y_2)A^+(y') A^+(y)A^+(y_1) \right]\rangle \,, \label{eq:T4_centralcut}
\end{align}
for each cut diagram, each possessing different step functions $\Theta$ that reflect different orderings as well as different perturbative factors $\mathcal{H}_{\rm D}$. Their complete expressions are shown in the supplemental material.
   As in the single scattering case, we expand $\mathcal{H}_{\rm D}$ in inverse powers of $p_{\gamma\perp}^2$ and we find (up to next-to-leading order):
\begin{align}
    &\mathcal{H}_{\rm D}(p_\gamma,y,y',y_1,y_2) = 8 \alpha_s  H^{(2)}_{q+g\to \gamma + q} e^{i x P_A^+ (y^- - y'^-)}  \nonumber \\
    & \times  \delta^{(2)}(\yt-\ytone ) \delta^{(2)}(\ytC-\yttwo ) \delta^{(2)}(\ytone-\yttwo )  \nonumber \\
    & \times  \left[1 + \frac{\mathcal{D}_{X}}{\pgammat^2} \ (\partial_{ \ytone} \cdot \partial_{\yttwo}) \right] (\partial_{\yt} \cdot \partial_{\ytC}) + \mathcal{O}(1/\pgammat^8)
    \label{eq-hexp}
    \,.
\end{align}

As in the LT case, the derivatives $(\partial_{\yt} \cdot \partial_{\ytC})$ transform two gauge fields into strength field tensors. Then, the first term in the square bracket in Eq.\,(\ref{eq-hexp}) 
contributes to the gauge link in LT gluon distribution function. The two additional derivatives $(\partial_{\ytone} \cdot \partial_{\yttwo})$, on the second term in the square bracket, promote the two remaining gauge fields into strength field tensors, thus providing the genuine $\rm{T4}$ contribution for double (and triple-single interference) scattering. The final result matches exactly the twist-4 result in Eq.\,(\ref{eq-t4}), including all the longitudinal phase factors, and the corresponding derivative operators $\mathcal{D}_X$. Thus, by including sub-eikonal phases and the diagrams that are responsible for the LPM effect in the finite-$x$ region, we finally achieve an exact matching between the CGC and HT. 

\vspace{0.5cm}
\noindent{\it Summary}.- We proved for the first time the consistency between CGC and collinear factorization formalism of twist expansion at up to twist-4 level for direct photon production in $pA$ collisions.  We clarify explicitly that the naive collinear expansion of CGC in terms of multiple scattering reproduces the leading twist result in the small-$x$ limit, while only recovering part of the complete result at twist-4. We emphasize two important missing ingredients in CGC that lead to the mismatch, i.e. sub-eikonal phases and diagrams related to the LPM interference, both of which are important at finite-$x$. Including these two missing ingredients in CGC, we show the exact matching to collinear factorization at leading twist in the dilute region and twist-4 in the relatively dense region,
\begin{align}
      \left. E_{\gamma} \frac{{\rm d}\sigma^{{\rm CGC_{sub}}}}{{\rm d}^3 \boldsymbol{p_{\gamma}}}\right|_{p_{\gamma\perp}> Q_s}
        = E_{\gamma} \frac{{\rm d}\sigma^{{\rm LT}}}{{\rm d}^3 \boldsymbol{p_{\gamma}}} +E_{\gamma} \frac{{\rm d}\sigma^{{\rm T4}}}{{\rm d}^3 \boldsymbol{p_{\gamma}}}+\cdots.
\end{align}

The methodology developed in this paper can be easily extended to any other processes, such as single inclusive hadron production in $pA$ collisions, and dijet production in deep inelastic scattering, as long as the CGC factorization is valid. Therefore, one can take full advantage of these processes to calculate their perturbative hard parts using our approach, and then map out the phase diagram from dilute to dense regions from existing RHIC and LHC data, and future measurements at Electron Ion Colliders \cite{AbdulKhalek:2021gbh,EicCWP,Anderle:2021wcy}.
We thus expect a very broad application of our new framework in $eA$ and $pA$ collisions, which can provide robust theoretical input for searching for signatures of gluon saturation. 

This work is supported by the NSFC under Grants Nos.~12022512, 12035007, 11890714 and 1935007, by the Guangdong Major Project of Basic and Applied Basic Research No.~2020B0301030008 (H.X.), by the US NSF Grant No.~PHY-1945471 (Z.K., F.S.) and
No. OAC-2004571 within the X-SCAPE Collaboration (F.S., X.W.), by the U.S. DOE under Contract No.~DE-AC02-05CH11231, and within the framework of the SURGE Collaboration.



\newpage
\widetext
\clearpage
\begin{center}

\textbf{\large Supplemental Material}

\end{center}

In this supplemental material, we list the full expressions for twist-4 and CGC results with sub-eikonal corrections. The detailed derivation of these results will be shown in a forthcoming companion paper \cite{Fu:2023xxx}. 

The complete twist-4 contribution contains not only the double scattering, but also the interference between single and triple scattering. The final result at twist-4 can be written as the following compact form
\bea
E_{\gamma}\frac{\der^3\sigma^{\rm T4}}{\der^3 \boldsymbol{p_{\gamma}}} =& f_{q/p}\otimes \mathcal{D}_{X}T_{gg} \otimes H^{(4)}_{q+gg\to \gamma+q}
\\ \nonumber
=&\int {\rm d}x_q f_{q/p}(x_q)  \frac{4\pi^2e_q^2\alpha_{\rm em}\alpha_s^2}{N_c^2} \frac{\left[1 + (1-\xi)^2 \right]}{\boldsymbol{p_{\gamma\perp}}^{\ 6}}  \Big[ \mathcal{D}_{X} T_{X}(x_1,x_2,x_3) \Big]_{x_1=x,x_2=x_3=0},
\eea
with
\begin{equation}
    T_{X}(x_1,x_2,x_3)= \int\frac{ {\rm d} y^- {\rm d} y_1^-}{2\pi} \frac{ {\rm d} y_2^-}{2\pi} e^{ix_1P_A^+ y^-}  e^{ix_2P_A^+ (y_1^- - y_2^-)}  e^{ix_3 P_A^+ y_2^-}
         \frac{1}{ P_A^+ }  F^4_X \Theta_X .
\end{equation}
Notice that, for the convenience of listing the full results, we use slightly different notation in the supplemental material for the twist-4 matrix element $T_{gg}$, where the subscript $gg$ is suppressed due to the reason that we only consider multiple scatterings with gluons in nuclei in this paper. Instead, we use $T_{X=A,B}$ with $A$ and $B$ stand for different cut and initial/final state multiple scattering, respectively. In particular, $A=C,L,R$ stand for central-cut (double scattering), left-cut (single-triple interference) and right-cut (triple-single interference), respectively. And $B=I,F,IF,FI$ represent multiple scattering for initial state, final state, initial-final interference and final-initial interferences, respectively. The detailed expressions for  $\mathcal{D}_{X}$, $F^4_{X}$ and $\Theta_X$ are presented in Table.~\ref{ht-tbl}.

Now, we list the hard factors before power expansion in the CGC with sub-eikonal corrections. For single scattering $\mathcal{H}_{\rm S}$ in Eq. (9) in the main text,
\begin{align}
    \mathcal{H}_{\rm S}(p_\gamma;y,y') = \frac{2 e_q^2 \alpha_{ \mathrm{em}}  \alpha_s}{\pi N_c} \left[  1+ (1-\xi)^2 \right] \xi^2 e^{i x P_A^+ (y^- - y'^-)} \int \frac{\der^2 \lt } {(2\pi)^2}  \frac{e^{-i \lt \cdot (\yt-\yt')} \lt^2}{(\xi \lt -\pgammat)^2 \pgammat^2} .
\end{align}

For double scattering as well as the interference between single-triple interference, i.e. $\mathcal{H}_{\rm D}$ in Eq. (11) in the main text,
\begin{align}
    \mathcal{H}_{\rm D}(p_\gamma;y,y',y_1,y_2) &= \frac{8 e_q^2 \alpha_{ \mathrm{em}}  \alpha_s^2}{N_c} \left[1 + (1-\xi)^2 \right]  \int \frac{\der^2 \Lt}{(2\pi)^2} \int \frac{\der^2 \lt}{(2\pi)^2} \int \frac{\der^2 \lt'}{(2\pi)^2} \nonumber \\
    & \times  e^{i (\Lt -\lt') \cdot \yttwo} e^{i \lt'  \cdot \yt'} e^{-i \lt  \cdot \yt} e^{-i (\Lt -\lt) \cdot \ytone} \mathcal{N}_{X},
\end{align}
where the subscript in $\mathcal{N}_{X}$ possesses the same meaning as that in $T_X$.

For central-cut, the corresponding $\mathcal{N}_{X}$ is 
\begin{align}
    \mathcal{N}_{\rm C,I} &= \left\{\frac{\left[ \xi \ltCL{\alpha}  - (\xi \LtL{\alpha}  -\pgammatL{\alpha}) \right]}{\left[\xi\ltC -( \xi \Lt - \pgammat ) \right]^2} + \frac{\left[ \xi \LtL{\alpha} - \pgammatL{\alpha} \right]}{\left[ \xi \Lt -\pgammat \right]^2}  \right\} \left\{\frac{\left[ \xi \lt^{\alpha}  - (\xi \LtU{\alpha}  -\pgammatU{\alpha}) \right]}{\left[\xi\lt -( \xi \Lt - \pgammat ) \right]^2} + \frac{\left[ \xi \LtU{\alpha} - \pgammatU{\alpha} \right]}{\left[ \xi \Lt -\pgammat \right]^2}  \right\} \nonumber \\
    &\times e^{-i \left[ \frac{\xi(1-\xi)(\Lt -\ltC)^2}{\pgammat^2} \right]x P_A^+ y_2^-} e^{-i \left[\frac{ \xi (\Lt-\pgammat)^2 + (1-\xi)\pgammat^2 - \xi(1-\xi) (\Lt -\ltC)^2}{\pgammat^2}\right] x P_A^+ y'^-}  \nonumber \\
    & \times e^{i \left[\frac{ \xi (\Lt-\pgammat)^2 + (1-\xi)\pgammat^2 - \xi(1-\xi)(\Lt -\lt)^2}{\pgammat^2}\right]x P_A^+ y^-}  e^{i \left[\frac{ \xi(1-\xi)(\Lt -\lt)^2}{\pgammat^2} \right] x P_A^+ y_1^-} \,.
\end{align}

\begin{align}
    \mathcal{N}_{\rm C,F} &= \left\{\frac{(\xi \ltCL{\alpha} -\pgammatL{\alpha})}{\left( \xi \ltC -\pgammat\right)^2}+\frac{\pgammatL{\alpha}}{\pgammat^2}  \right \} \left\{\frac{(\xi \ltU{\alpha} -\pgammatU{\alpha})}{\left( \xi \lt -\pgammat\right)^2}+\frac{\pgammatU{\alpha}}{\pgammat^2}  \right \}  \nonumber \\
    & \times  e^{-i \left[\frac{ (1-\xi)\pgammat^2 +\xi(\ltC-\pgammat)^2}{\pgammat^2} \right] x P_A^+ y'^-} e^{-i\left[\frac{\xi((\Lt-\pgammat)^2 - (\ltC - \pgammat)^2)}{\pgammat^2} \right] x P_A^+ y_2^- } \nonumber \\
    & \times  e^{i\left[\frac{\xi((\Lt-\pgammat)^2 - (\lt - \pgammat)^2)}{\pgammat^2} \right] x P_A^+ y_1^- } e^{i \left[\frac{ (1-\xi)\pgammat^2 +\xi(\lt-\pgammat)^2}{\pgammat^2} \right] x P_A^+ y^-} \,.
\end{align}


\begin{table}[H]
\renewcommand{\arraystretch}{1.45}
    \begin{center}
         \caption{$\mathcal{D}_{X}$ and $T_{X}$ for different types of scattering with different cuts}     \label{ht-tbl}
          \resizebox{\hsize}{!}{
        \begin{tabular}{|l|c|l|}
            \hline
            \multirow{9}{*}{\makecell[{}{p{1.4cm}}]{Initial\\ state\\ scattering }} & \multirow{3}{*}{\makecell[{}{p{1.1cm}}]{Central\\Cut}} & $\mathcal{D}_{C,I}= \xi^4 x^2\frac{\partial^2}{\partial x_1^2}-3\xi^4 x \frac{\partial}{\partial x_1}+(1-\xi)\xi^3 x \frac{\partial}{\partial x_2} +4\xi^4$  \\
            \cline{3-3}
            &    & $F^4_{C,I}=   \langle P_A|F^{+\alpha}(y_2^-) F^{+\beta}(0^-) F^{+}_{\ \ \beta}(y^-) F^+_{\ \  \alpha}(y_1^-) |P_A\rangle $  \\
            \cline{3-3}
            &    & $\Theta_{C,I}= \theta(y^- - y_1^-)\theta(-y_2^-) $  \\
            \cline{2-3}

            & \multirow{3}{*}{\makecell[{}{p{1.1cm}}]{Left\\Cut}}  & $\mathcal{D}_{L,I}= (1-\xi)\xi^3 x \frac{\partial}{\partial x_2}$  \\
            \cline{3-3}
            &    & $F^4_{L,I}=   \langle P_A|F^{+\alpha}(y_2^-) F^+_{\ \  \alpha}(y_1^-) F^{+\beta}(0^-) F^{+}_{\ \ \beta}(y^-)  |P_A\rangle $   \\
            \cline{3-3}
            &    & $\Theta_{L,I}=  \theta(y_1^- - y_2^-)\theta(-y_1^-) $   \\
            \cline{2-3}

            & \multirow{3}{*}{\makecell[{}{p{1.1cm}}]{Right\\Cut}}  & $\mathcal{D}_{R,I}= (1-\xi)\xi^3 x \frac{\partial}{\partial x_2}$  \\
            \cline{3-3}
            &    & $F^4_{R,I}= \langle P_A|F^{+\beta}(0^-) F^{+}_{\ \ \beta}(y^-) F^{+\alpha}(y_2^-)  F^+_{\ \  \alpha}(y_1^-) |P_A\rangle $   \\
            \cline{3-3}
            &    & $\Theta_{R,I}= \theta(y^- - y_2^-)\theta(y_2^- - y_1^-) $   \\
            \cline{2-3}

            \hline\hline

            \multirow{9}{*}{\makecell[{}{p{1.4cm}}]{Final\\ state\\ scattering }} & \multirow{3}{*}{\makecell[{}{p{1.1cm}}]{Central\\Cut}} & $\mathcal{D}_{C,F}= \xi^4 x^2\frac{\partial^2}{\partial x_2^2}+\xi^3 x \frac{\partial}{\partial x_2}$  \\
            \cline{3-3}
            &    & $F^4_{C,F}= \langle P_A| F^{+\beta}(0^-) F^{+\alpha}(y_2^-) F^+_{\ \  \alpha}(y_1^-) F^{+}_{\ \ \beta}(y^-) |P_A\rangle $  \\
            \cline{3-3}
            &    & $\Theta_{C,F}=  \theta(y_1^- - y^-)\theta(y_2^-) $  \\
            \cline{2-3}

            & \multirow{3}{*}{\makecell[{}{p{1.1cm}}]{Left\\Cut}}  & $\mathcal{D}_{L,F}= -\xi^4 x^2\frac{\partial^2}{\partial x_2^2} - \xi^3 x \frac{\partial}{\partial x_2}$  \\
            \cline{3-3}
            &    & $F^4_{L,F}= \langle P_A| F^{+\beta}(0^-) F^{+\alpha}(y_2^-) F^+_{\ \  \alpha}(y_1^-) F^{+}_{\ \ \beta}(y^-) |P_A\rangle $   \\
            \cline{3-3}
            &    & $\Theta_{L,F}=  \theta(y_1^- - y_2^-)\theta(y_2^-) $   \\
            \cline{2-3}

            & \multirow{3}{*}{\makecell[{}{p{1.1cm}}]{Right\\Cut}}  & $\mathcal{D}_{R,F}=  -\xi^4 x^2\frac{\partial^2}{\partial x_2^2} - \xi^3 x \frac{\partial}{\partial x_2} $  \\
            \cline{3-3}
            &    & $F^4_{R,F}=  \langle P_A| F^{+\beta}(0^-) F^{+\alpha}(y_2^-) F^+_{\ \  \alpha}(y_1^-) F^{+}_{\ \ \beta}(y^-) |P_A\rangle $   \\
            \cline{3-3}
            &    & $\Theta_{R,F}=  \theta(y_2^- - y_1^-)\theta(y_1^- - y^-) $   \\
            \cline{2-3}

            \hline\hline

            \multirow{6}{*}{\makecell[{}{p{1.7cm}}]{Initial\\ Final\\ interference }} & \multirow{3}{*}{\makecell[{}{p{1.1cm}}]{Central\\Cut}} & $\mathcal{D}_{C,IF}= \xi^4 x^2(\frac{\partial}{\partial x_1} - \frac{\partial}{\partial x_3})^2 -\xi^4 x(\frac{\partial}{\partial x_1} - \frac{\partial}{\partial x_3}) + (1-\xi)\xi^3 x \frac{\partial}{\partial x_2} $  \\
            \cline{3-3}
            &    & $F^4_{C,IF}=   \langle P_A| F^{+\beta}(0^-) F^{+\alpha}(y_2^-) F^{+}_{\ \ \beta}(y^-) F^+_{\ \  \alpha}(y_1^-) |P_A\rangle $  \\
            \cline{3-3}
            &    & $\Theta_{C,IF}= \ \theta(y^- - y_1^-)\theta(y_2^-) $  \\
            \cline{2-3}

            & \multirow{3}{*}{\makecell[{}{p{1.1cm}}]{Right\\Cut}}  & $\mathcal{D}_{R,IF}=  -\xi^4 x^2(\frac{\partial}{\partial x_1} - \frac{\partial}{\partial x_3})^2 + \xi^4 x(\frac{\partial}{\partial x_1} - \frac{\partial}{\partial x_3}) - (1-\xi)\xi^3 x \frac{\partial}{\partial x_2} $  \\
            \cline{3-3}
            &    & $T_{R,IF}=   \langle P_A|  F^{+\beta}(0^-) F^{+\alpha}(y_2^-) F^{+}_{\ \ \beta}(y^-) F^+_{\ \  \alpha}(y_1^-) |P_A\rangle $  \\
            \cline{3-3}
            &    & $\Theta_{R,IF}=  \theta(y_2^- - y^-)\theta(y^- - y_1^-) $  \\
            \cline{2-3}

            \hline\hline

            \multirow{6}{*}{\makecell[{}{p{1.7cm}}]{Final\\ Initial\\ interference }} & \multirow{3}{*}{\makecell[{}{p{1.1cm}}]{Central\\Cut}} & $\mathcal{D}_{C,FI}=  \xi^4 x^2(\frac{\partial}{\partial x_2} + \frac{\partial}{\partial x_3})^2   + (1-2\xi)\xi^3 x \frac{\partial}{\partial x_2} - \xi^4 x \frac{\partial}{\partial x_3} $  \\
            \cline{3-3}
            &    & $F^4_{C,FI}=   \langle P_A| F^{+\alpha}(y_2^-) F^{+\beta}(0^-)  F^+_{\ \  \alpha}(y_1^-) F^{+}_{\ \ \beta}(y^-) |P_A\rangle $  \\
            \cline{3-3}
            &    & $\Theta_{C,FI}=  \theta(y_1^- - y^-) \theta(-y_2^-) $  \\
            \cline{2-3}

            & \multirow{3}{*}{\makecell[{}{p{1.1cm}}]{Left\\Cut}}  & $\mathcal{D}_{L,FI}= -\xi^4 x^2(\frac{\partial}{\partial x_2} + \frac{\partial}{\partial x_3})^2   - (1-2\xi)\xi^3 x \frac{\partial}{\partial x_2} + \xi^4 x \frac{\partial}{\partial x_3} $  \\
            \cline{3-3}
            &    & $F^4_{L,FI}=  \langle P_A|  F^{+\alpha}(y_2^-) F^{+\beta}(0^-)  F^+_{\ \  \alpha}(y_1^-) F^{+}_{\ \ \beta}(y^-) |P_A\rangle$  \\
            \cline{3-3}
            &    & $\Theta_{L,FI}=  \theta(y_1^-) \theta(-y_2^-)$  \\
            \cline{2-3}

            \hline

        \end{tabular}
     }
    \end{center}
\end{table}

\begin{align}
    \mathcal{N}_{\rm C,IF} &= \left\{\frac{(\xi \ltCL{\alpha} -\pgammatL{\alpha})}{\left( \xi \ltC -\pgammat\right)^2}+\frac{\pgammatL{\alpha}}{\pgammat^2}  \right \} \left\{\frac{\left[ \xi \lt^{\alpha}  - (\xi \LtU{\alpha}  -\pgammatU{\alpha}) \right]}{\left[\xi\lt -( \xi \Lt - \pgammat ) \right]^2} + \frac{\left[ \xi \LtU{\alpha} - \pgammatU{\alpha} \right]}{\left[ \xi \Lt -\pgammat \right]^2}  \right\} \nonumber \\
    &\times e^{-i \left[\frac{ (1-\xi)\pgammat^2 +\xi(\ltC-\pgammat)^2}{\pgammat^2} \right] x P_A^+ y'^-} e^{-i\left[\frac{\xi((\Lt-\pgammat)^2 - (\ltC - \pgammat)^2)}{\pgammat^2} \right] x P_A^+ y_2^- }  \nonumber \\
    & \times  e^{i \left[\frac{ \xi (\Lt-\pgammat)^2 + (1-\xi)\pgammat^2 - \xi(1-\xi)(\Lt -\lt)^2}{\pgammat^2}\right]x P_A^+ y^-} e^{i \left[\frac{ \xi(1-\xi)(\Lt -\lt)^2}{\pgammat^2} \right] x P_A^+ y_1^-} \,.
\end{align}
\begin{align}
    \mathcal{N}_{\rm C,FI} &= \left\{\frac{\left[ \xi \ltCL{\alpha}  - (\xi \LtL{\alpha}  -\pgammatL{\alpha}) \right]}{\left[\xi\ltC -( \xi \Lt - \pgammat ) \right]^2} + \frac{\left[ \xi \LtL{\alpha} - \pgammatL{\alpha} \right]}{\left[ \xi \Lt -\pgammat \right]^2}  \right\} \left\{\frac{(\xi \ltU{\alpha} -\pgammatU{\alpha})}{\left( \xi \lt -\pgammat\right)^2}+\frac{\pgammatU{\alpha}}{\pgammat^2}  \right \} \nonumber \\
    & \times e^{-i \left[ \frac{\xi(1-\xi)(\Lt -\ltC)^2}{\pgammat^2} \right]x P_A^+ y_2^-}  e^{-i \left[\frac{ \xi (\Lt-\pgammat)^2 + (1-\xi)\pgammat^2 - \xi(1-\xi) (\Lt -\ltC)^2}{\pgammat^2}\right] x P_A^+ y'^-} \nonumber \\
    & \times e^{i\left[\frac{\xi((\Lt-\pgammat)^2 - (\lt - \pgammat)^2)}{\pgammat^2} \right] x P_A^+ y_1^- } e^{i \left[\frac{ (1-\xi)\pgammat^2 +\xi(\lt-\pgammat)^2}{\pgammat^2} \right] x P_A^+ y^-}  \,.
\end{align}
For left Cut,
\begin{align}
    \mathcal{N}_{\rm L,I} &= -\left\{\frac{\left[ \xi  \ltCL{\alpha} -\pgammatL{\alpha} \right] }{\left[ \xi  \ltC - \pgammat \right]^2} + \frac{ \pgammatL{\alpha} }{\pgammat^2} \right\} \left\{\frac{\left[ \xi \lt^{\alpha} - ( \xi \ltCU{\alpha} -\pgammatU{\alpha}) \right]}{\left[\xi\lt -( \xi \ltC - \pgammat ) \right]^2} + \frac{\left[ \xi \ltCU{\alpha} - \pgammatU{\alpha} \right]}{\left[ \xi \ltC - \pgammat \right]^2}  \right\}  
    \nonumber \\
    & \times e^{-i x P_A^+ y'^-} e^{i \left[ \frac{\xi (\ltC-\pgammat)^2 +(1-\xi)\pgammat^2 -\xi(1-\xi)(\ltC -\lt)^2}{\pgammat^2}\right] x P^+_A y^-} \nonumber \\
    & \times e^{i \left[ \frac{\xi(1-\xi)((\ltC -\lt)^2 - \Lt^2)}{\pgammat^2} \right] x P^+_A y_2^- } e^{i \xi (1-\xi)\Lt^2 x P^+_A y_1^-} \,.
\end{align}
\begin{align}
    \mathcal{N}_{\rm L,F} &= -\left\{\frac{\left[ \xi  \ltCL{\alpha}-\pgammatL{\alpha} \right] }{\left[ \xi  \ltC- \pgammat \right]^2} + \frac{ \pgammatL{\alpha} }{\pgammat^2} \right\} \left\{\frac{(\xi \ltU{\alpha} -\pgammatU{\alpha})}{\left( \xi \lt -\pgammat\right)^2}+\frac{\pgammatU{\alpha}}{\pgammat^2}  \right \}  \nonumber \\
    & \times e^{-i x P_A^+ y'^-} e^{i \left[ \frac{\xi( (\ltC-\pgammat)^2 - (\pgammat-\lt -\Lt)^2 )}{\pgammat^2} \right] x P^+_A y_2^-} \nonumber \\
    &\times e^{i\left[ \frac{\xi((\pgammat - \lt -\Lt)^2 - (\pgammat-\lt)^2 )}{\pgammat^2}  \right] x P^+_A  y_1^- } e^{i \left[ \frac{(1-\xi)\pgammat^2 + \xi (\pgammat-\lt)^2 }{\pgammat^2} \right] x P^+_A  y^-} \,.
\end{align}

\begin{align}
    \mathcal{N}_{\rm L,IF} &= -\left\{\frac{\left[ \xi  \ltCL{\alpha}-\pgammatL{\alpha} \right] }{\left[ \xi  \ltC- \pgammat \right]^2} + \frac{ \pgammatL{\alpha} }{\pgammat^2} \right\} \left\{ \frac{\left[ \xi \lt^{\alpha}  - (\pgammatU{\alpha} -\xi \LtU{\alpha}) \right] }{  \left[ \xi\lt -(\pgammat -\xi \Lt)\right]^2} + \frac{\left[  \pgammatU{\alpha} - \xi \LtU{\alpha}\right] }{  \left[ \pgammat -\xi \Lt  \right]^2} \right\}   \nonumber \\
    & \times e^{-i x P_A^+ y'^-} e^{i \frac{1}{2k^-}\left[ \frac{(\ltC-\pgammat)^2}{1-\xi}  - \frac{(\pgammat -\lt -\Lt)^2}{1-\xi}\right] y_2^-} \nonumber \\
    & \times e^{i \frac{1}{2k^-}\left[ \frac{(\pgammat -\lt -\Lt)^2}{1-\xi} + \frac{\pgammat^2}{\xi} - \Lt^2 \right] y^- } e^{i \frac{1}{2k^-} \Lt^2 y_1^-} \,.
\end{align}
For right Cut,
\begin{align}
    \mathcal{N}_{\rm R,I} &= -\left\{\frac{\left[ \xi \lt^{\alpha} - ( \xi \ltCU{\alpha} -\pgammatU{\alpha}) \right]}{\left[\xi\lt -( \xi \ltC - \pgammat ) \right]^2} + \frac{\left[ \xi \ltCU{\alpha} - \pgammatU{\alpha} \right]}{\left[ \xi \ltC - \pgammat \right]^2}  \right\} \left\{\frac{\left[ \xi  \ltCL{\alpha}-\pgammatL{\alpha} \right] }{\left[ \xi  \ltC- \pgammat \right]^2} + \frac{ \pgammatL{\alpha} }{\pgammat^2} \right\}
    \nonumber \\
    & \times e^{-i \xi (1-\xi)\Lt^2 x P^+_A y_2^-} e^{-i \left[ \frac{\xi(1-\xi)((\ltC -\lt)^2 - \Lt^2)}{\pgammat^2} \right] x P^+_A y_1^- } \nonumber \\
    & \times e^{-i \left[ \frac{\xi (\ltC-\pgammat)^2 +(1-\xi)\pgammat^2 -\xi(1-\xi)(\ltC -\lt)^2}{\pgammat^2}\right] x P^+_A y'^-}  e^{i x P_A^+ y^-} \,.
\end{align}

\begin{align}
    \mathcal{N}_{\rm R,F} &= -\left\{\frac{(\xi \ltU{\alpha} -\pgammatU{\alpha})}{\left( \xi \lt -\pgammat\right)^2}+\frac{\pgammatU{\alpha}}{\pgammat^2}  \right \} \left\{\frac{\left[ \xi  \ltCL{\alpha}-\pgammatL{\alpha} \right] }{\left[ \xi  \ltC- \pgammat \right]^2} + \frac{ \pgammatL{\alpha} }{\pgammat^2} \right\} \nonumber \\
    & \times  e^{-i \left[ \frac{(1-\xi)\pgammat^2 + \xi (\pgammat-\lt)^2 }{\pgammat^2} \right] x P^+_A  y'^-} e^{-i\left[ \frac{\xi((\pgammat - \lt -\Lt)^2 - (\pgammat-\lt)^2 )}{\pgammat^2}  \right] x P^+_A  y_2^- } \nonumber \\
    & \times e^{-i \left[ \frac{\xi( (\ltC-\pgammat)^2 - (\pgammat-\lt -\Lt)^2 )}{\pgammat^2} \right] x P^+_A y_1^-} e^{i x P_A^+ y^-} \,.
\end{align}

\begin{align}
    \mathcal{N}_{\rm R,FI} &= -\left\{ \frac{\left[ \xi \lt^{\alpha}  - (\pgammatU{\alpha} -\xi \LtU{\alpha}) \right] }{  \left[ \xi\lt -(\pgammat -\xi \Lt)\right]^2} + \frac{\left[  \pgammatU{\alpha} - \xi \LtU{\alpha}\right] }{  \left[ \pgammat -\xi \Lt  \right]^2} \right\} \left\{\frac{\left[ \xi  \ltCL{\alpha}-\pgammatL{\alpha} \right] }{\left[ \xi  \ltC- \pgammat \right]^2} + \frac{ \pgammatL{\alpha} }{\pgammat^2} \right\}  \nonumber \\
    & \times e^{-i \frac{1}{2k^-} \Lt^2 y_2^-} e^{-i \frac{1}{2k^-}\left[ \frac{(\pgammat -\lt -\Lt)^2}{1-\xi} + \frac{\pgammat^2}{\xi} - \Lt^2 \right] y'^- }  \nonumber \\
    & \times e^{-i \frac{1}{2k^-}\left[ \frac{(\ltC-\pgammat)^2}{1-\xi}  - \frac{(\pgammat -\lt -\Lt)^2}{1-\xi}\right] y_1^-}  e^{i x P_A^+ y^-} \,.
\end{align}


\end{document}